# Exploring the Integration of Differential Privacy in Cybersecurity Analytics: Balancing Data Utility and Privacy in Threat Intelligence

Brahim Khalil Sedraoui[1,] Abdelmadjid Benmachiche[1,] Amina Makhlouf[1,] and Chaouki Chemam[1]

[1] *University of Chadli Bendjedid, Faculty of Sciences & Technology, El Tarf, Algeria*

**Abstract**

To resolve the acute problem of privacy protection and guarantee that data can be used in the context of threat intelligence, this paper considers the implementation of Differential Privacy (DP) in cybersecurity analytics. DP, which is a sound mathematical framework, ensures privacy by adding a controlled noise to data outputs and thus avoids sensitive information disclosure even with auxiliary datasets. The use of DP in Security Information and Event Management (SIEM) systems is highlighted, and it can be seen that DP has the capability to protect event log and threat data analysis without interfering with the analytical efficiency. The utility versus privacy trade-offs linked to the maximization of the epsilon parameter, which is one of the critical components of DP mechanisms, is pointed out. The article shows the transformative power of DP in promoting safe sharing of data and joint threat intelligence through real-world systems and case studies. Finally, this paper makes DP one of the key strategies to improve privacy-preserving analytics in the field of cybersecurity.

## 1. Introduction to Differential Privacy and Cybersecurity Analytics

Differential privacy (DP) has become a widely accepted framework for preserving the privacy of individual entries in a dataset when statistical analyses of the data are published, including data generated and analyzed within large-scale digital platforms such as online learning systems, MOOCs [1], [2], [3], and Open Classrooms [4], [5], [6]. DP was introduced in the computer science and statistics literature in 2006 for providing formal guarantees about the privacy of the individuals described by a dataset, while allowing for useful generalizations to be made and shared about that dataset [7].

As a generalization, the privacy of the individuals in a dataset is protected by noise addition. The fundamental DP framework describes a general process for sharing and analyzing a dataset were given a dataset $D$, a function $f$, and an $\varepsilon > 0$, for all neighboring datasets $D$ and $D0$ (i.e., datasets that differ by at most one individual), the output of the function $f$ satisfies: $\Pr[p(f(D))] \leq e^{\varepsilon} \Pr[p(f(D0))]$ (1) for any potential outcome p. That is, the two probabilities cannot differ by a factor larger than $e^{\varepsilon}$ [8]. Three core properties characterize DP:

(1) an output for an analysis must not compromise privacy;

(2) a crucial component of an analysis is a privacy-sensitive randomization process, where some noise is added to the output;

(3) the privacy guarantees do not depend on how well the adversary knows the dataset, or how many auxiliary sources of information the adversary has.

### 1.1. Definition and Principles of DP

DP is formally defined as a property of a subjective randomized algorithm $\mathcal{M}$ with domain $\mathbb{N}|\mathcal{X}|$ is $(\epsilon, \delta)$ - deferentially private if for all S ⊆ Range($\mathcal{M}$) and for all $x, y \epsilon \mathbb{N}|\mathcal{X}|$ such that $||x - y||1 \leq 1$ will follow: (Formula), where the probability space is over the coin flips of the mechanism $\mathcal{M}$. DP is

inspired by the idea of maximizing the potential of public-interest data to help, while at the same time minimizing its corresponding risk to individual lives [7].

Core concepts in DP are its privacy guarantees. According to its definition, even if an adversary knows nearly all other databases used by the same mechanism $\mathcal{M}$ and observes the output y, they should not be able to learn anything significant about the specific database, such as whether it is included or not. The term 'almost' refers to the fact that there are always some events with non-zero probability for which the given mechanism $\mathcal{M}$ ensures privacy, but the two databases may differ in their output probability distributions, potentially leading to significant risks [9].

## 2. The Need for DP in Cybersecurity Analytics

Recent incidents involving data breaches targeting high-profile government agencies and businesses are indicative of a rising trend of advanced persistent threats (APTs). Threat-hunting teams play a crucial role in analytical activities involving the collection and examination of threat intelligence data concerning APT groups, malicious software (malware) [10], and cyber-attacks [11]. Traditionally, such data was shared as a wholesome file, but a burgeoning trend of sharing observational attributes by using Interactive Data Sharing Language (IDSL) language has emerged recently. Since this data contains sensitive intelligence, it is imperative to thoroughly examine its privacy-preserving methodology before employing it for collaborative intelligence analysis [7]

While academics have explored privacy-preserving data (PPD) techniques in academia and across numerous domains amply, there is no concurrent endeavor in the cybersecurity domain. By cybersecurity, reference is made to all activities surrounding the collection and analysis of threat data to build a shield against malicious cyber activity. Since such data sharing could suddenly leak sensitive intelligence, it fosters a motive to explore strong privacy-preserving techniques like DP. DP offers a strong mathematical promise of privacy, making it superior to many earlier PPD techniques [8]. Most existing DP data-sharing mechanisms are not congruous with IDSL-based analytics. Consequently, exploring such compatibility issues within analytics necessitates cross-laying prior knowledge from both analytics and DP perspectives, followed by suggestions to bridge the currently existing dichotomy therein.

### 2.1. Challenges in Preserving Privacy in Threat Intelligence Data

Organizations often seek to leverage threat intelligence from other sources to strengthen their cybersecurity defenses [12]. However, sharing raw threat intelligence data can expose sensitive information, thereby increasing the risk of exploitation by malicious actors. Prior to sharing, organizations must redress data privacy concerns, which may include obscuring sensitive information such as IP addresses, user IDs, and other unique identifiers that link back to an organization's internal environment [7].

Conventional measures for protecting privacy, such as data anonymization and obfuscation, may be inadequate. While anonymization can hide unique identifying attributes of the data, several studies have demonstrated that adversaries can still infer sensitive information through a combination of auxiliary knowledge and other identifying attributes (e.g., dates of birth, zip codes, etc.) [13]. Redacting specific values in the data may also be insufficient, as accurately guessing them may still lead to privacy violations. Through estimation approaches (e.g., using misspecified models), it is possible to make inferences about the redacted information. Additionally, shared datasets often have auxiliary datasets that can assist in inferring sensitive information (e.g., using a social network graph).

The notions of privacy have often focused on direct and specific privacy breaches, severely underestimating the information that can be leaked from the data. Therefore, there is a need for privacy-preserving mechanisms that provide rigorous privacy guarantees even when background knowledge is employed, and privacy is not compromised by the combination of several other pieces of information.

## 3. Applications of DP in Cybersecurity

Security Information and Event Management (SIEM) systems are widely adopted to centrally analyze security data generated by a variety of devices and technologies, including autonomous mobile robots [14], [15], [16], [17], speech recognition systems [18], [19], [20], [21], and navigation systems [22], where sensitive operational data must be protected while enabling meaningful analysis. Typically, this involves deploying an agent on the device that collects and prepares the data before transmitting it to a SIEM infrastructure. In this context, DP can be employed on the agents collecting events by adding noise to aggregation queries that analyze the events before forwarding the information to the SIEM back end [23]. This mitigates the risk of exposing sensitive individual information while still making it possible to derive significant insights from the data.

The promising results achieved in this research work indicate that it is feasible to integrate various forms of DP with existing SIEM tools, while satisfying constraints on the utility of the data. Reducing the risk of breaching personal information would also allow organizations to share their logs without fear of penalties. This is especially relevant in cases where it is desirable to collectively analyze statistics of different organizations' logs, e.g., to assess general threat intelligence. More generally, this would enable the distributed deployment of agents, which brings scalability gains.

### 3.1. Integrating DP with SIEM Systems for Secure Data Analysis

The surging frequency of cyber threats has led to the more rapid adoption SIEM systems, enabling enterprises to gather event data for analysis. An SIEM system is a unified process for managing system logs and security events that utilizes analysis and correlation to detect and alert the presence of aberrant activity [24]. The collected security data are stored in a database and provide the foundation for an analysis that is performed based on several views of the data, including user activity, access attempts, and file modification. However, the accessibility of sensitive security data in and outside an enterprise raises a serious concern regarding privacy and potential exposure. Cyber analysis is a powerful and extensible analytical method for big data that tracks queries and data quality and provides a mechanism for unconstrained analytics [7]. Big data fuels improvements in learning and comprehending data. Moreover, big data cannot be anonymized and released, as anonymity cannot provide strict privacy guarantees when the data is released. This has led to a new approach to big data protection based on differential privacy (DP), which provides privacy guarantees based on the inclusion or exclusion of any single record in the database.

Nevertheless, a mechanism for safely analyzing security data without jeopardizing privacy has not yet gained traction. Meanwhile, DP mechanisms often require extensive query constraints and can reduce the utility of the data being analyzed, which precludes complicated queries and exhaustive coverage in data analysis. Consequently, an SIEM system is integrated with DP for safe and efficient data analysis, where security data is fenced by a PPD analyzer enforcing DP and ensuring that queries and updates to the data comply with the DP policy [25]. A fence query algorithm makes optimal use of DP-compliant OMIN operations, and a two-move approach for a reported mechanism ensures that sensitive security data are not exposed in any direct request to the SIEM system.

## 4. Balancing Data Utility and Privacy in Threat Intelligence

While there is no single, universally accepted solution to the problem of balancing privacy and utility in any particular application, social scientists and policy analysts need to better understand the complexity of decision-making in relation to trade-offs. There is a robust discussion of privacy-utility trade-offs in the literature on DP, although such discussions are on the technical side, discussing the mathematical framework of trade-offs and privacy-loss budgets in terms of information theory [26]. Threat intelligence functionalists should empirically examine the balancing of privacy and security instrumental within the actual practices of cybersecurity analysts. From the perspective of cybersecurity analysts, data utility is understood as useability for cybersecurity analysis and intelligence production, flowing from consideration of both usefulness for a given analysis and veracity, quality, and precision. Threat intelligence is data or information that can be used to counter threats and vulnerabilities posed to the assets and information systems of organizations of all kinds, including digital educational platforms such as Intelligent Tutoring Systems [27], [28]. Within the trade-off, two areas of consideration arise – the privacy preserving techniques put into play and the impact upon data by implementing such techniques. From here, the most actionable goal is a trajectory towards optimal epsilon, or ε parameter. Indeed, this sensitivity region may be narrowed around a choice of ε for differentially-private mechanisms. $ε's$ elasticity entails a privacy/utility trade-off [29]. If ε is large, the mechanisms have better utility from the accessibility of greater amounts of the data set. But privacy guarantees are weaker, as more of the unaltered data is fed to the algorithm—there is a non-linear relationship between ε and the probability of violation. Conversely, if ε is small, mechanisms provide greater privacy against external attacks, but lower data utility.

### 4.1. Optimizing Epsilon Parameter for Trade-offs

To balance the trade-off between data privacy and data usefulness, the epsilon parameter in DP must be optimized. A mechanism's level of privacy is measured by a positive value called epsilon (ε); the lower the epsilon, the more stringent the privacy assurances, but the higher the data utility loss. On the other hand, a greater epsilon provides lower privacy but better data usefulness [30]. The application's objectives and context must be taken into account in order to maximize Epsilon. This entails figuring out the epsilon value for cybersecurity analytics that protects people's privacy while optimizing the value of threat intelligence data. Adjusting epsilon entails testing with various values, assessing the effect on data utility (for example, using metrics like completeness or correctness), and making sure the privacy constraints as outlined by ethical and legal norms are fulfilled [26]. Performing synthetic tests with different epsilon values and examining the impact of noise generated by the Laplace or other processes on the data's analytical value are common steps in the optimization process. Organizations can select an epsilon that satisfies their privacy and utility requirements by weighing the trade-offs, finding a balance that preserves sensitive data while enabling efficient data use for security research[29].

## 5. Case Studies and Practical Implementations

This section will detail real-world case studies and practical implementations that illustrate the use of DP in cybersecurity analytics. By exploring tangible examples and experiences where DP has been effectively employed, valuable insights will be gained into the concrete applications of DP within the cybersecurity domain [12]. This study will examine how DP has been integrated into cybersecurity analytics practices, the challenges faced, and the impact of this integration. Through these case studies, the potential of DP in protecting data from various applications, including Recommendation Systems [31], while still enabling meaningful analysis will be demonstrated.

Cybersecurity analytics involves the examination of data across networks, servers, devices, and users to identify and mitigate risks [32]. Organizations are increasingly adopting data-driven analytics as their primary strategy for dealing with cyber threats. However, in order to effectively analyze data for the purpose of cybersecurity investigations and to power detection algorithms, security analytics systems often rely on large volumes of sensitive and personal data [33]. Data breaches of security analytics systems could have far-reaching implications for organizations, as well as harms for individuals. To mitigate privacy risks in such systems that analyze sensitive data, DP has been explored as a solution [7]. DP allows data sharing with strong privacy guarantees, ensuring that the outputs of a query do not significantly differ with or without an individual's data in the data set.

## 5.1. Real-world Examples of DP in Cybersecurity Analytics

In cybersecurity, DP is essential for doing data analysis while preserving individual privacy. By adding noise to data, DP helps SIEM systems detect threats without disclosing private information [7]. By combining data in a manner that makes it impossible to identify particular entities, DP also makes it easier to share threat intelligence in a safe collaborative manner. Adding noise to statistical summaries in malware research helps create detection signatures and preserve individual data samples [34].

DP helps with user and network activity monitoring by protecting the privacy of behavior analysis while detecting any dangers [3], [35]. In incident response, DP assists with data analysis without jeopardizing the privacy of individuals [34]. Machine learning models may be developed on sensitive data while maintaining privacy thanks to the combination of DP with deep learning [36]. Managing computing costs, maintaining regulatory compliance, and striking a balance between privacy and value are some of the difficulties. Notwithstanding these difficulties, DP's contribution to cybersecurity analytics is essential for preserving privacy and strengthening security protocols.

## 6. Evaluating the Effectiveness of DP in Cybersecurity Analytics

The effectiveness of DP in preserving privacy is evaluated using metrics and criteria from the perspective of threat intelligence analysis. Given raw data such as IP addresses used for cyber-attacks, reports on detected malware, and variants of collected malware, DP is applied to address general queries asked on the dataset, such as the number of queries in a specified time range. The relationship between the query and the privacy parameter is shown to reduce the sensitivity of the query counts, thereby preserving privacy. The output of these queries is appropriately perturbed to satisfy DP.

The effectiveness of DP is assessed under two aspects: (i) the likelihood of preserving privacy as intended after applying DP and (ii) the likelihood of success in gaining public knowledge from the output after applying DP. The average DP is used as the posterior privacy guarantee [37], and DP can be considered to preserve privacy. On the other hand, knowledge of the query and database provides a B-attack: the attacker estimates the effect of the database on the output of the query, and if the estimate is above a threshold, S will be deduced. A sufficient condition is formulated on the query count's privacy parameter such that the probability of success of the B-attack is below a certain threshold.

## 6.1. Metrics for Assessing Privacy Preservation

Different approaches have been implemented to assess the preservation of privacy in regard to DP solutions. There are two main groups of metrics and parameters to assess the preservation of privacy in datasets. Metrics belong mainly to one of two categories: entropy based and re-identification based [38]. Both groups of metrics compute privacy scores from 0-1, but their interpretation is reversed since higher scores demonstrate better privacy in the first group and in the second group lower scores demonstrate better privacy.

Another function of the privacy related metrics is to compute the amount of protection that a privacy model or a particular anonymization method grants to the dataset. For a few popular privacy models, or classes of them, these metrics have been computed and they are published in articles. These metrics have been independently implemented and incorporated into the assessment framework to test the protection provided to the datasets. The efficacy of the metrics has been examined on a few benchmark datasets for DP preserving data publishing solutions [39].

## 7. Future Directions and Emerging Trends in DP

While the previously discussed techniques of DP in the Cyber Security domain address most of the critical emerging problems and challenges, there are many advancements and developments in the domain of such DP techniques that have a huge potential to help in shaping the future of this DP domain. Some of the most exciting and promising advancements/modifications/evolutions in DP techniques are discussed in this section which includes but is not limited to Adaptive Sample Size Selection Methods, Adaptive Mechanisms, Hybrid Mechanisms, Differential Privacy-Oriented Systems and Frameworks, Privacy-Preserving Hybrid Systems of Cyber Security, Applications of Deep Learning Framework in Protecting Cyber Security via DP, and Emerging Threats and Defense Mechanisms for DP in Cyber Security [7].

Different advancements and developments related to DP techniques can be implemented for malware data, as malware remains one of the most significant threats in cybersecurity. Numerous advanced modifications in malware threat analytics have been proposed in the literature, primarily relying on evolving computational intelligence (CI) techniques and parallel processing or multi-core frameworks, with recent studies also incorporating transformer-based deep learning models to capture complex temporal and contextual patterns in malware behavior [40]. In the future, DP could be used in combination with these advanced malware threat models, as well as federated learning paradigms that enable collaborative model training across distributed organizations without sharing raw data, to increase performance efficiency while ensuring secure data disclosure. Furthermore, blockchain-based infrastructures can be leveraged to provide transparency, integrity, and auditability for privacy-preserving malware intelligence sharing and model update exchanges [41]. However, there is an increasing number of quantifiable attacks targeting DP mechanisms, making it necessary to identify such attack vectors and develop robust security design principles against them. Researchers in the cybersecurity domain should therefore focus on establishing resilient DP models that remain secure against adaptive and collusion-based attacks. Moreover, the growing attempts to bypass DP by circumventing privacy filters highlight the need for robust cybersecurity filtering and verification mechanisms, potentially supported by blockchain-based auditing, which may open new research directions and enable more secure and reliable comparative results [42].

### 7.1. Potential Advancements in DP Techniques for Cybersecurity

The cybersecurity sector recognizes the need to address privacy issues arising from the desire to share valuable data (such as threat intelligence), and uses various anonymization techniques (such as data suppression or noise addition transformation) [42]. This academic focus will probably widen DP to handle other domains and dataset types, such as non-numerical data and relational datasets like medical datasets [43]. So far, most DP works explore classic time, instance, and attribute level protections. Other dataset aspects offer exciting avenues for DP protection, such as complex graph type or unstructured High-Dimensional datasets. That will open a chance to explore the impacts of advanced DP protection types on cybersecurity analytics tasks. A noteworthy example would be the incorporation of Local-DP data protection [7] approaches within sensors.

Risk evaluation and privacy guarantees are two key aspects yet to be further explored on DP for all analyzed analytics tasks while uncertainty propagation and analysis in data analytic processes, as this task matrix shows many inconceivable effects. Another exciting opportunity lies in the study of the impacts of various attack strategies on DP-type guarantees to assess the privacy of the datasets. Tackling this opportunity can be done in conjunction with one of the former two avenues, which is to examine implications of conducting analyses on different strategies on the same data sets and compare results.

## 8. Conclusion and Key Takeaways

Threat intelligence (TI) is an essential aspect of cybersecurity but leads to significant privacy issues that can be explained by the fact that sensitive information about an organization may be spread. Traditional threat intelligence practices often do not provide sufficient protection to privacy. In turn, the current literature has shifted focus to incorporating the concept of DP in threat intelligence systems to reduce information disclosure. The available literature is used to evaluate the use of DP algorithms, outline the privacy-utility trade-offs, and measure the protective efficacy provided. The analysis of large intrusion detection logs has been presented with a DP-based mechanism that has been developed with the aim of preserving analytic value and limiting privacy threats [44]. However, there are still substantive issues, such as how to select appropriate privacy-utility trade-offs, reduce the costs of secure data-sharing: many of the solutions that still exist poorly characterize or justify their privacy-utility balance to be used in practice [7].

A proper privacy-preserving TI analysis can unlock the potential of sharing logs from different organizations, increasing the visibility of emerging threats. This would make it harder for attackers to perform reconnaissance and have a better understanding of the defended environment. It would offer realistic defense solutions for low-budget companies and prevent the widespread of malicious tools. On the other hand, it is important to mention that adopting differential privacy mechanisms does not eliminate the need for already proposed TI privacy-aware systems. There are many ways that private information can be leaked, and differential privacy guarantees do not provide protection against every possibility [45]. Thus, it is a collective approach that should be adopted in conjunction with other privacy protection mechanisms.